\documentclass[conference,letterpaper]{IEEEtran}

\usepackage[utf8]{inputenc} 
\usepackage[T1]{fontenc}
\usepackage{url}
\usepackage{ifthen}
\usepackage{cite}
\usepackage[cmex10]{amsmath} 
\usepackage{amsfonts}

\begin{document}
\title{Non-Asymptotic Fundamental Limits of Guessing Subject to Distortion} 


\author{\IEEEauthorblockN{Shota Saito and Toshiyasu Matsushima}
\IEEEauthorblockA{Department of Applied Mathematics, Waseda University, Tokyo, 169-8555 JAPAN\\ 
E-mail: shota@aoni.waseda.jp and toshimat@waseda.jp}
}


\maketitle

\begin{abstract}
This paper investigates the problem of guessing subject to distortion, which was introduced by Arikan and Merhav.
While the primary concern of the previous study was asymptotic analysis, our primary concern is non-asymptotic analysis.
We prove non-asymptotic achievability and converse bounds of the moment of the number of guesses without side information (resp. with side information) by using a quantity based on the R\'enyi entropy (resp. the Arimoto-R\'enyi conditional entropy).
Also, we introduce an error probability and show similar results.
Further, from our bounds, we derive a single-letter characterization of the asymptotic exponent of guessing moment for a stationary memoryless source.
\end{abstract}


\section{Introduction}
\subsection{Prior Work and Research Motivation}
The problem of guessing is one of the research topics in Shannon theory.
Roughly speaking, this problem is described as a game of guessing a realization of a discrete random variable $X$ by asking questions of the form ``Is $X$ equal to $x$ ?''.
Let $G(x)$ be the number of guesses required by a guessing strategy when $X = x$.
Then, one of the interesting questions is ``What is the fundamental limit of the $\rho$-th moment\footnote{$\rho>0$ is a given parameter.} of the number of guesses $\mathbb{E}[G(X)^\rho]$?''.
Several previous studies for this question are summarized as follows:\footnote{Note that the studies of the problem of guessing are not limited to those described here. See, e.g., the introduction of \cite{Sason} and references therein.}
\begin{itemize}
\item[1)] Arikan \cite{Arikanlossless} derived the {\it non-asymptotic} achievability and converse bounds of the $\rho$-th moment of the number of guesses $\mathbb{E}[G(X)^\rho]$ by using the {\it R\'enyi entropy}. 
He also investigated the problem of guessing with side information and characterized the {\it non-asymptotic} achievability and converse bounds of the $\rho$-th moment of the number of guesses by using the {\it Arimoto-R\'enyi conditional entropy}.
Further, the recent study by Sason and Verd\'u \cite{Sason} provided the improved non-asymptotic bounds by using the {\it R\'enyi entropy} and the {\it Arimoto-R\'enyi conditional entropy}.
\item[2)] Kuzuoka \cite{Kuzuoka} investigated a guessing problem allowing errors, in which a guesser may give up guessing and declare an error. 
He derived the {\it non-asymptotic} achievability and converse bounds of the $\rho$-th moment of the number of guesses by using the {\it conditional smooth R\'enyi entropy} when side information is available.
\item[3)] Arikan and Merhav \cite{Arikan} introduced a distortion and investigated the problem of guessing subject to distortion. 
Their primary concern was the {\it asymptotic} analysis. 
They characterized the fundamental limit of the $\rho$-th moment of the number of guesses without side information (resp.\ with side information) by using the {\it rate-distortion function and relative entropy} (resp.\ the {\it conditional rate-distortion function and relative entropy}) under a stationary memoryless source when blocklength goes to infinity.
\end{itemize}

The above previous studies raise the following question, which motivated us to carry out this research.
\begin{itemize}
\item[($\clubsuit$)] Regarding the guessing subject to distortion, is it possible to derive {\it non-asymptotic} achievability and converse bounds by using a quantity related to the {\it R\'enyi entropy}?
\end{itemize}
This paper gives a positive answer to the question ($\clubsuit$).

\subsection{Contributions}
For the problem of guessing subject to distortion, our first contribution is to prove non-asymptotic achievability and converse bounds of the $\rho$-th moment of the number of guesses without side information (resp.\ with side information) by using a quantity based on the R\'enyi entropy (resp.\ the Arimoto-R\'enyi conditional entropy).
To derive the results, we utilize the techniques developed in \cite{Saito18}.
When we consider the special case in the problem setup, our achievability bound reduces to the bound in \cite{Arikanlossless} and our converse bound is slightly tighter than the results in \cite{Arikanlossless}.

Our second contribution is to introduce an error probability to the problem of guessing subject to distortion and derive non-asymptotic achievability and converse bounds of the $\rho$-th moment of the number of guesses. In this case, these bounds are also characterized by the quantity based on the R\'enyi entropy (resp.\ the Arimoto-R\'enyi conditional entropy) when there is no side information (resp.\ there is side information).

Our third contribution is to derive a single-letter characterization of the $\rho$-th moment of the number of guesses for a stationary memoryless source when blocklength goes to infinity and $\rho \downarrow 0$.
Our result in the case without errors coincides with the result of Arikan and Merhav \cite{Arikan}.
Further, our result in the case of allowing errors indicates the benefit of allowing a positive error probability.

\subsection{Organization of the Paper}
Section \ref{setup} describes the problem setup. 
Section \ref{PreviousStudy} reviews the related previous results in \cite{Arikan}.
Section \ref{Main} shows our results.
Proofs of one-shot achievability and converse bounds are in Sec.\ \ref{Proof}.
Finally, in Sec.\ \ref{Extension}, we introduce an error probability and show our results in the case allowing errors.

\section{Problem Setup} \label{setup}

\newtheorem{theorem}{Theorem}
\newtheorem{condi}{Condition}
\newtheorem{defi}{Definition}
\newtheorem{lem}{Lemma}
\newtheorem{cor}{Corollary}
\newtheorem{proof}{Proof}
\newtheorem{remark}{Remark}
\newcommand{\argmax}{\mathop{\rm arg~max}\limits}
\newcommand{\argmin}{\mathop{\rm arg~min}\limits}

This section reviews the setup of the problem of guessing subject to distortion described in \cite{Arikan}.
Let ${\cal X}$ be a finite set called a source alphabet and $\hat{{\cal X}}$ be a finite set called a reproduction alphabet.
Let $X$ be a random variable 
taking a value in ${\cal X}$ and $x$ be a realization of $X$.
The probability distribution of $X$ is denoted as $P_{X}$.
A distortion measure $d$ is defined as
$
d : {\cal X} \times \hat{{\cal X}} \rightarrow [0, +\infty).
$
We assume that for ever $x \in {\cal X}$, there exsits $\hat{x} \in \hat{{\cal X}}$ such that $d (x, \hat{x}) = 0$.

A {\it $D$-admissible guessing strategy} is defined as follows.
\begin{defi}
For $D \geq 0$, a $D$-admissible guessing strategy with respect to $P_X$ is a set
$
{\cal G} = \{ \hat{x} (1), \hat{x} (2),  \cdots \} \subset \hat{{\cal X}}
$
such that
\begin{align} 
\mathbb{P} [d (X, \hat{x} (j)) \leq D ~~ {\rm for ~ some } ~ j ] =1.
\end{align} 
In the following, the elements in a guessing strategy ${\cal G}$ are referred to as {\it guessing codewords}.
\end{defi}

By using a particular $D$-admissible guessing strategy ${\cal G} = \{ \hat{x} (1), \hat{x} (2),  \cdots \}$, a guesser asks ``Does $x$ satisfy $d(x, \hat{x}(i)) \leq D$?'' successively until the answer is ``Yes.''
In this situation, we are interested in the number of guesses required by a particular guessing strategy when $X=x$.
This quantity is defined as a {\it guessing function} as follows.
\begin{defi}
The guessing function $G(\cdot)$ induced by a $D$-admissible guessing strategy ${\cal G}$ is the function that maps each $x \in {\cal X}$ into a positive integer, which is the index $j$ of the first guessing codeword $\hat{x} (j) \in {\cal G}$ such that $d (x, \hat{x} (j)) \leq D$.
\end{defi}

The quantity we investigate is the {\it $\rho$-th moment of the number of guesses} for a given parameter $\rho > 0$:
$
\frac{1}{\rho} \log \mathbb{E}[G (X)^\rho ],
$
where all logarithms are of base 2 throughout this paper.

As an extension of the above setup, we formulate the guessing problem with side information, in which a guesser observes side information $y$ taking a value in a finite set ${\cal Y}$.
Let $P_{XY}$ denote a joint probability distribution of $X$ and $Y$ and $P_{X|Y}$ denote a conditional probability distribution of $X$ given $Y$.
Then, a {\it $D$-admissible guessing strategy with side information} is defined as follows.
\begin{defi}
For $D \geq 0$, a $D$-admissible guessing strategy with side information is a set ${\cal G}_s = \{ {\cal G}(y) : y \in {\cal Y} \}$, such that for every $y \in {\cal Y}$,
$
{\cal G}(y) = \{ \hat{x}_y (1), \hat{x}_y (2),  \cdots \}
$
is a $D$-admissible guessing strategy with respect to $P_{X | Y=y}$.
\end{defi}

Next, we define a {\it guessing function with side information}.
\begin{defi}
The guessing function with side information $G(x|y)$ induced by ${\cal G}_s$ is the function that maps each $(x, y) \in {\cal X} \times {\cal Y}$ into a positive integer, which is the index $j$ of the first guessing codeword $\hat{x}_y (j) \in {\cal G}(y)$ such that $d (x, \hat{x}_y (j)) \leq D$.
\end{defi}

For a given parameter $\rho > 0$, the quantity we investigate is the {\it $\rho$-th moment of the number of guesses with side information}:
$
\frac{1}{\rho} \log \mathbb{E}[G (X|Y)^\rho ].
$

\section{Previous Study} \label{PreviousStudy}
Let $Q_X$ and  $Q_{XY}$ be a probability distribution on ${\cal X}$ and ${\cal X} \times {\cal Y}$, respectively.
For $D \geq 0$, let $R(D, Q_X)$ and $R(D, Q_{XY})$ be the rate-distortion function and the conditional rate-distortion function, respectively\footnote{More precisely, $R(D, Q_X) :=  \inf I(X; \hat{X})$, where $I(X; \hat{X})$ is the mutual information and the infimum is taken over all conditional probability distribution $P_{\hat{X} |X}$ such that $\mathbb{E} [d(X,\hat{X})] = \sum_{x, \hat{x}} Q_X (x) P_{\hat{X} |X} (\hat{x} | x) d(x, \hat{x}) \leq D$. Regarding the definition of $R(D, Q_{XY})$, see (80) in \cite{Arikan}.}.
Also, let ${\rm KL}(\cdot || \cdot)$ denote the relative entropy.
Using these quantities, Arikan and Merhav \cite{Arikan} characterized the asymptotic fundamental limits of the $\rho$-th moment of the number of guesses for a stationary memoryless source.

\begin{theorem}[\cite{Arikan}] \label{Prev}
Suppose that a distortion measure 
$
d_n : {\cal X}^n \times \hat{{\cal X}}^n \rightarrow [0, +\infty)
$
satisfies 
$
d_n(x^n, \hat{x}^n) = \sum_{i=1}^{n} d(x_i, \hat{x}_i)
$
and a source is stationary and memoryless.
Then, for any $D \geq 0$ and $\rho > 0$, we have
\begin{align} 
\lim_{n \to \infty} & \frac{1}{n \rho} \inf_{\substack{{\cal G}_n }} \log \mathbb{E}[G_{n}(X^n)^\rho ] \notag \\
& = \sup_{\substack{Q_X}}\left(R(D, Q_X)- \frac{1}{\rho} {\rm KL}(Q_X || P_X) \right), \label{pre1} \\
\lim_{n \to \infty} & \frac{1}{n \rho} \inf_{\substack{{\cal G}_{s, n} }} \log \mathbb{E}[G_{n}(X^n | Y^n)^\rho ] \notag \\
& = \sup_{\substack{Q_{XY}}}\left(R(D, Q_{XY})- \frac{1}{\rho} {\rm KL}(Q_{XY} || P_{XY}) \right), \label{pre2}
\end{align}
where 
the infimum in (\ref{pre1}) (resp.\ (\ref{pre2})) is taken over all $D$-admissible guessing strategies ${\cal G}_n$ (resp.\ all $D$-admissible guessing strategies with side information ${\cal G}_{s, n}$),
$G_{n}(\cdot)$ in (\ref{pre1}) (resp.\ $G_{n}(\cdot | \cdot)$ in (\ref{pre2})) is a guessing function induced by ${\cal G}_n$ (resp.\ ${\cal G}_{s, n}$), 
and the supremum in (\ref{pre1}) (resp.\ (\ref{pre2})) is taken over all probability distribution $Q_X$ on ${\cal X}$ (resp.\ $Q_{XY}$ on ${\cal X} \times {\cal Y}$).
\end{theorem}

\section{Main Results} \label{Main}
\subsection{One-Shot Achievability and Converse Bounds} \label{MainNonAsymptotic}
For $\alpha \in (0,1) \cup (1,\infty)$, the R\'enyi entropy $H_{\alpha}(X)$ \cite{Renyi} and the Arimoto-R\'enyi conditional entropy $H_{\alpha}(X | Y) $ \cite{Arimoto77} are defined as 
\begin{align*}
& H_{\alpha}(X) := \frac{1}{1 - \alpha} \log \sum_{x \in {\cal X}} [P_{X}(x)]^{\alpha}, \\
& H_{\alpha}(X | Y) := \frac{\alpha}{1 - \alpha} \log \sum_{y \in {\cal Y}} P_{Y}(y)
\left( \sum_{x \in {\cal X}} [P_{X | Y}(x | y)]^{\alpha}\right)^{\frac{1}{\alpha}}.
\end{align*}

Based on these quantities, we introduce new quantities, which play an important role in producing our results.
\begin{defi}
For $D \geq 0$ and $\alpha \in (0,1) \cup (1,\infty)$, $\mathbb{H}^{D}_{\alpha}(X)$ and $\mathbb{H}^{D}_{\alpha}(X | Y)$ are defined as
\begin{align}
&\mathbb{H}^{D}_{\alpha}(X) := \inf_{\substack{P_{\hat{X}|X} : \\
\mathbb{P} [ d(X, \hat{X}) \leq D ] = 1 }} H_{\alpha}(\hat{X}), \\
&\mathbb{H}^{D}_{\alpha}(X|Y) := \inf_{\substack{P_{\hat{X}|X,Y} : \\
\mathbb{P} [ d(X, \hat{X}) \leq D ] = 1 }} H_{\alpha}(\hat{X}|Y).   
\end{align}
\end{defi}

Now, the next theorem shows the achievability bounds.
\begin{theorem} \label{OneShotAchievability}
For any $D \geq 0$ and $\rho > 0$, there exists a $D$-admissible guessing strategy ${\cal G}^*$ such that
\begin{align}
\frac{1}{\rho} \log \mathbb{E}[G^{*}(X)^\rho ]  \leq \mathbb{H}^{D}_{\frac{1}{1+\rho}}(X), \label{OneShotA1}
\end{align}
where $G^{*} (\cdot)$ is the guessing function induced by ${\cal G}^*$.
Also, there exists a $D$-admissible guessing strategy with side information ${\cal G}^{*}_{s}$ such that
\begin{align}
\frac{1}{\rho} \log \mathbb{E}[G^{*} (X | Y)^\rho ]  \leq \mathbb{H}^{D}_{\frac{1}{1+\rho}}(X | Y), \label{OneShotA2}
\end{align}
where $G^{*} (\cdot | \cdot)$ is the guessing function induced by ${\cal G}^{*}_{s}$.
\end{theorem}

\begin{remark} 
Let us consider the case where $D=0$, ${\cal X} = \hat{{\cal X}}$, $d(x, \hat{x}) = 0$ for $x = \hat{x}$, and $d(x, \hat{x}) = 1$ for $x \neq \hat{x}$.
In this case, the bounds (\ref{OneShotA1}) and (\ref{OneShotA2}) reduce to 
\begin{align}
\frac{1}{\rho} \log \mathbb{E}[G^{*}(X)^\rho ] & \leq H_{\frac{1}{1+\rho}}(X), \\
\frac{1}{\rho} \log \mathbb{E}[G^{*}(X | Y)^\rho ] & \leq H_{\frac{1}{1+\rho}}(X | Y),
\end{align}
respectively.
These bounds coincide with the results in \cite{Arikanlossless}.
\end{remark} 

The next theorem shows the converse bounds.
\begin{theorem} \label{OneShotConverse}
Fix $D \geq 0$ and $\rho > 0$.
For any $D$-admissible guessing strategy ${\cal G}$, we have
\begin{align} 
\frac{1}{\rho}  \log \mathbb{E}[G(X)^\rho ] \geq \mathbb{H}^{D}_{\frac{1}{1+\rho}}(X) - \log \log (1+ \min\{|{\cal X}|, |\hat{\cal X}| \}), \label{OneShotC1}
\end{align}
where $G (\cdot)$ is the guessing function induced by ${\cal G}$ and $| \cdot |$ denotes the cardinality of a set.
Also, for any $D$-admissible guessing strategy with side information ${\cal G}_s$, we have
\begin{align} 
\frac{1}{\rho} & \log \mathbb{E}[G (X|Y)^\rho ] \notag \\
& \geq \mathbb{H}^{D}_{\frac{1}{1+\rho}}(X|Y) - \log \log (1+ \min\{|{\cal X}|, |\hat{{\cal X}}| \}), \label{OneShotC2}
\end{align}
where $G (\cdot | \cdot)$ is the guessing function induced by ${\cal G}_s$.
\end{theorem}

\begin{remark} 
Let us consider the case where $D=0$, ${\cal X} = \hat{{\cal X}}$, $d(x, \hat{x}) = 0$ for $x = \hat{x}$, and $d(x, \hat{x}) = 1$ for $x \neq \hat{x}$.
In this case, the bounds (\ref{OneShotC1}) and (\ref{OneShotC2}) reduce to 
\begin{align}
\hspace{-3mm} \frac{1}{\rho} \log \mathbb{E}[G(X)^\rho ] & \geq H_{\frac{1}{1+\rho}}(X) - \log \log (1+|{\cal X}|), \\
\hspace{-3mm} \frac{1}{\rho} \log \mathbb{E}[G (X | Y)^\rho ] & \geq H_{\frac{1}{1+\rho}}(X | Y) - \log \log (1+|{\cal X}|),
\end{align}
respectively.
These bounds are slightly tighter than the results in \cite{Arikanlossless}.
\end{remark} 

\begin{IEEEproof}
The proof of (\ref{OneShotA1}) (resp.\ (\ref{OneShotC1})) is in Sec.\ \ref{ProofOneShotAchievability} (resp.\ Sec.\ \ref{ProofOneShotConverse}).
The bound with side information (\ref{OneShotA2}) (resp.\ (\ref{OneShotC2})) can be readily shown by the argument proceeded in the proof of (\ref{OneShotA1})  (resp.\ (\ref{OneShotC1})).
More precisely, we can derive (\ref{OneShotA2}) (resp.\  (\ref{OneShotC2})) by replacing $P_X$ with $P_{X|Y=y}$ in the proof of Sec.\ \ref{ProofOneShotAchievability} (resp.\ Sec.\ \ref{ProofOneShotConverse}).
\end{IEEEproof}

\subsection{Asymptotics for a Stationary Memoryless Source}
This section investigates the bounds (\ref{OneShotA1}) and (\ref{OneShotC1}) when a stationary memoryless source is assumed.
In particular, we consider the special case $\rho \downarrow 0$ and drive a single-letter characterization of the fundamental limit of the $\rho$-th moment of the number of guesses when blocklength $n$ goes to infinity.

First, we can immediately obtain the next corollaries by using the same argument which is used to prove (\ref{OneShotA1}) and (\ref{OneShotC1}).

\begin{cor} \label{BlocklengthAchievability}
For any $D \geq 0$, $ \rho > 0$, and blocklength $n \in \mathbb{N}$, there exists a $D$-admissible guessing strategy ${\cal G}^{*}_{n}$ such that
\begin{align}
\frac{1}{n \rho} \log \mathbb{E}[G^{*}_{n}(X^n)^\rho ]  \leq \frac{1}{n} \mathbb{H}^{D}_{\frac{1}{1+\rho}}(X^n),
\end{align}
where $G^{*}_{n} (\cdot)$ is the guessing function induced by ${\cal G}^{*}_{n}$.
\end{cor}

\begin{cor} \label{BlocklengthConverse}
Fix $D \geq 0$, $\rho > 0$, and blocklength $n \in \mathbb{N}$.
For any $D$-admissible guessing strategy ${\cal G}_n$, we have
\begin{align} 
& \frac{1}{n \rho} \log \mathbb{E}[G_{n}(X^n)^\rho ] \notag \\
& \geq \frac{1}{n} \mathbb{H}^{D}_{\frac{1}{1+\rho}}(X^n) - \frac{1}{n} \log \log (1+ \min\{|{\cal X}^n|, |\hat{{\cal X}}^n| \}),
\end{align}
where $G_n (\cdot)$ is the guessing function induced by ${\cal G}_n$.
\end{cor}

Next, regarding the term $\frac{1}{n} \mathbb{H}^{D}_{1/(1+\rho)}(X^n)$, our previous study \cite{Saito18} has proved the following result for a stationary memoryless source with a distortion measure
$
d_n : {\cal X}^n \times \hat{{\cal X}}^n \rightarrow [0, +\infty)
$
such that
$
d_n(x^n, \hat{x}^n) = \sum_{i=1}^{n} d(x_i, \hat{x}_i)
$:\footnote{More precisely, (\ref{HR}) is a special version of (46) in \cite{Saito18}.}
\begin{align} 
\lim_{\rho \downarrow 0} \frac{1}{n} \mathbb{H}^{D}_{\frac{1}{1+\rho}}(X^n) = R(D, P_X) + O \left( \frac{\log n}{n} \right). \label{HR}
\end{align}

Combining Corollaries \ref{BlocklengthAchievability}, \ref{BlocklengthConverse}, and (\ref{HR}), we obtain the next theorem.
\begin{theorem} 
Suppose that a distortion measure
$
d_n : {\cal X}^n \times \hat{{\cal X}}^n \rightarrow [0, +\infty)
$
satisfies 
$
d_n(x^n, \hat{x}^n) = \sum_{i=1}^{n} d(x_i, \hat{x}_i)
$
and a source is stationary and memoryless.
Then, we have
\begin{align} 
\lim_{n \to \infty} \lim_{\rho \downarrow 0} \frac{1}{n\rho} \inf_{\substack{{\cal G}_n }} \log \mathbb{E}[G_{n}(X^n)^\rho ] = R(D, P_X), \label{asymptotic}
\end{align}
where the infimum is taken over all $D$-admissible guessing strategies ${\cal G}_n$ and $G_n(\cdot)$ is a guessing function induced by ${\cal G}_n$.
\end{theorem}

\begin{remark} 
By examining the proof of (\ref{pre1}) and using Proposition 1 in \cite{Arikan}, we obtain
\begin{align} 
\lim_{n \to \infty} \lim_{\rho \downarrow 0} \frac{1}{n\rho} \inf_{\substack{{\cal G}_n }} \log \mathbb{E}[G_{n}(X^n)^\rho ] = R(D, P_X),
\end{align}
which coincides with (\ref{asymptotic}).
\end{remark} 

\section{Proof of One-Shot Bounds} \label{Proof}
\subsection{Proof of Achievability Bound (\ref{OneShotA1})} \label{ProofOneShotAchievability}
First, some notations are defined.
\begin{itemize}
\item For any $\hat{x} \in \hat{{\cal X}}$ and $D \geq 0$, a distortion $D$-ball centered at $\hat{x}$ is defined as
$
{\cal B}_D (\hat{x}) := \{ x \in {\cal X} : d(x,\hat{x}) \leq D \}. \label{BD}
$

\item 
We define  $\hat{x}^{*}(i)$ ($i = 1, 2, \cdots$) by the following procedure.
Let $\hat{x}^{*}(1)$ be defined as
\begin{align}
\hat{x}^{*}(1) := \argmax_{\hat{x} \in \hat{{\cal X}}} \mathbb{P} [ X \in {\cal B}_{D} (\hat{x}) ].
\end{align}
For $i = 2, 3, \cdots, \tau$ (suppose that the procedure stops at $\tau$), let $\hat{x}^{*}(i)$ be defined as
\begin{align} 
\hspace{-3mm} \hat{x}^{*}(i) = \argmax_{\hat{x} \in \hat{{\cal X}}} \mathbb{P} \left [ X \in {\cal B}_{D} (\hat{x}) \setminus \bigcup_{j=1}^{i-1} {\cal B}_{D} (\hat{x}^{*}(j)) \right ]
\end{align}

\item We define ${\cal A}_D(\hat{x}^{*}(1))$ as
$
{\cal A}_D(\hat{x}^{*}(1)) = {\cal B}_{D} (\hat{x}^{*}(1)), \label{SetA1} 
$
and for $i=2, 3, \ldots, \tau$, we define ${\cal A}_D(\hat{x}^{*}(i))$ by
\begin{align}
{\cal A}_D(\hat{x}^{*}(i)) &= {\cal B}_{D} (\hat{x}^{*}(i)) \setminus \bigcup_{j=1}^{i-1} {\cal B}_{D}(\hat{x}^{*}(j)). \label{SetA}
\end{align}
From the definition, we have
\begin{align}
& \bigcup_{j=1}^{i} {\cal A}_{D} (\hat{x}^{*}(j)) = \bigcup_{j=1}^{i} {\cal B}_{D} (\hat{x}^{*}(j)) \quad (\forall i \geq 1), \label{A1} \\
& {\cal A}_{D} (\hat{x}^{*}(i)) \cap {\cal A}_{D} (\hat{x}^{*}(j)) =  \emptyset \quad (\forall i \neq j), \label{A2} \\
& {\cal X} = \bigcup_{j=1}^{\tau} {\cal A}_{D}(\hat{x}^{*}(j)), \\
 & \mathbb{P} [ X \in {\cal A}_{D} (\hat{x}^{*}(1)) ] \geq \mathbb{P} [ X \in {\cal A}_{D} (\hat{x}^{*}(2)) ] \notag \\ 
 &~~ \geq \cdots \geq \mathbb{P} [ X \in {\cal A}_{D} (\hat{x}^{*}(\tau)) ].
\end{align}
\end{itemize}

Now, we construct the guessing strategy ${\cal G}^*$ as
$
{\cal G}^* = \{ \hat{x}^{*}(1), \hat{x}^{*}(2), \cdots, \hat{x}^{*}(\tau) \},
$
and let $G^{*} (\cdot)$ be the guessing function induced by ${\cal G}^*$.
Then, from the definition of $\hat{x}^{*}(1), \hat{x}^{*}(2), \cdots, \hat{x}^{*}(\tau)$, we see that ${\cal G}^*$ is a $D$-admissible guessing strategy with respect to $P_X$.  
Further, the $\rho$-th moment of the number of guesses for the guessing strategy ${\cal G}^*$ is evaluated as follows:
\begin{align} 
\frac{1}{\rho} & \log \mathbb{E}[G^{*}(X)^\rho ]  
 = \frac{1}{\rho} \log \sum_{x \in {\cal X}} P_X(x) G^{*}(x)^\rho \\
& \overset{(a)}{=} \frac{1}{\rho} \log \left( \sum_{i=1}^{\tau} \mathbb{P} [ X \in {\cal A}_{D} (\hat{x}^{*}(i)) ] i^{\rho} \right) \\
& \overset{(b)}{\leq}  \frac{1}{\rho} \log \left( \sum_{i=1}^{\tau} \mathbb{P} [ X \in {\cal A}_{D} (\hat{x}^{*}(i)) ] \right. \notag \\
& \hspace{16mm} \left. \times \left [ \sum_{j=1}^{\tau} \left (\frac{\mathbb{P} [ X \in {\cal A}_{D} (\hat{x}^{*}(j))]}{\mathbb{P} [ X \in {\cal A}_{D} (\hat{x}^{*}(i)) ]} \right )^{\frac{1}{1+\rho}} \right ]^{\rho} \right) \\
& = \frac{1+ \rho}{\rho} \log \sum_{k=1}^{\tau} [\mathbb{P} [ X \in {\cal A}_{D} (\hat{x}^{*}(k))]]^{\frac{1}{1+\rho}} \\
& \overset{(c)}{=} H_{\frac{1}{1+\rho}}(g(f(X))) 
\overset{(d)}{=} \mathbb{H}^{D}_{\frac{1}{1+\rho}}(X),
\end{align}
where
$(a)$ follows from the definition of ${\cal A}_{D} (\hat{x}^{*}(i))$,
$(b)$ is due to Lemma 3 in \cite{Saito18},
$(c)$ follows from the definition of the R\'enyi entropy and the variable-length lossy source code $(f, g)$ in Appendix \ref{code}, and
$(d)$ follows from (\ref{appendfg}) in Appendix \ref{code}.

\subsection{Proof of Converse Bound (\ref{OneShotC1})} \label{ProofOneShotConverse}
Let ${\cal G}$ be a $D$-admissible guessing strategy with guessing function $G (\cdot)$ and let $p_i$ be
\begin{align} 
p_i := \sum_{x \in {\cal X}: G(x) = i} P_{X}(x).
\end{align}
Then, the $\rho$-th moment of the number of guesses of this guessing strategy is calculated as 
\begin{align} 
\frac{1}{\rho} \log \mathbb{E}[G(X)^\rho ]  
= \frac{1}{\rho} \log \sum_{i=1}^{t} p_{i} i^\rho \label{ConvLower}
\end{align}
for some $t \in \mathbb{N}$.
Next, let $q_i ~ (i=1, \ldots, t)$ be a rearrange version of $p_i ~ (i=1, \ldots, t)$ in decreasing order of its value, i.e., $q_i$ satisfies $q_1 \geq q_2 \geq \ldots \geq q_t$ and $q_j = p_k$ for some $j, k \in \{1,2, \ldots, t \}$.
Then, it is easy to see that
\begin{align} 
\sum_{i=1}^{t} p_{i} i^\rho \geq \sum_{i=1}^{t} q_{i} i^\rho. \label{PA}
\end{align}
Combining (\ref{ConvLower}) and (\ref{PA}), we have
\begin{align} 
\frac{1}{\rho} & \log \mathbb{E}[G(X)^\rho ] \geq \frac{1}{\rho} \log \sum_{i=1}^{t} q_i i^\rho \\
& \overset{(a)}{\geq} \frac{1}{\rho} \log \sum_{i=1}^{\tau} \mathbb{P} [ X \in {\cal A}_{D} (\hat{x}^{*}(i)) ] i^{\rho} \label{majo} \\
& \overset{(b)}{\geq} \frac{1}{\rho} \log \sum_{i=1}^{\tau} \mathbb{P} [ X \in {\cal A}_{D} (\hat{x}^{*}(i)) ] 2^{\rho \ell(g^{-1}(\hat{x}^{*}(i)))} \\
& \overset{(c)}{=} \frac{1}{\rho} \log \mathbb{E}[2^{\rho \ell(f(X))}] \\
& \overset{(d)}{\geq} \mathbb{H}^{D}_{\frac{1}{1+\rho}}(X) - \log \log (1+ \min\{|{\cal X}|, |\hat{{\cal X}}| \}),
\end{align}
where
$(a)$ is verified by using the notion of majorization (see Appendix \ref{majorization}),
$(b)$ follows from (\ref{codebound}) in Appendix \ref{code},
$(c)$ is due to the construction of the code $(f, g)$ shown in Appendix \ref{code}, and
$(d)$ follows from Lemma 2 in \cite{Saito18}. 

\section{Extension: guessing allowing errors} \label{Extension}
As an extension of the problem formulation described in Sec.\ \ref{setup}, we introduce an ``error probability'' and consider the following setup\footnote{Note that the error probability considered in this section is different from the error considered in \cite{Kuzuoka}.}.
Given a guessing strategy\footnote{Here, a guessing strategy is not necessarily a $D$-admissible guessing strategy.}
$
\mathfrak{G} = \{ \hat{x} (1), \hat{x} (2),  \cdots \} \subset \hat{{\cal X}},
$
the guessing function $\mathfrak{g}(\cdot)$ induced by a guessing strategy $\mathfrak{G}$ is defined as the function that maps each $x \in {\cal X}$ into a positive integer, which is the index $j$ of the first guessing codeword $\hat{x} (j) \in \mathfrak{G}$ such that $d (x, \hat{x} (j)) \leq D$.
If there is no such guessing codeword, the guessing function declares an error.
Hence, the error probability $P_e (\mathfrak{G})$ for a guessing strategy 
$
\mathfrak{G} = \{ \hat{x} (1), \hat{x} (2),  \cdots \}
$
is given by
\begin{align}
P_e (\mathfrak{G}) = \mathbb{P} [d (X, \hat{x} (j)) > D ~~ {\rm for ~ all } ~ j ].
\end{align}
Note that if a guessing strategy $\mathfrak{G}$ is a $D$-admissible guessing strategy, we have $P_e (\mathfrak{G})= 0$.

In this setup, the achievability and converse bounds are given as follows.

\begin{theorem} \label{ExOneShotAchievability}
For any $D \geq 0$, $\rho > 0$, and $\epsilon \in [0,1)$, there exists a guessing strategy $\mathfrak{G}^*$ such that $P_e (\mathfrak{G}^*) \leq \epsilon$ and 
\begin{align}
\frac{1}{\rho} \log \mathbb{E}[\mathfrak{g}^* (X)^\rho ]  \leq \mathbb{H}^{D, \epsilon}_{\frac{1}{1+\rho}}(X), 
\end{align}
where $\mathfrak{g}^{*} (\cdot)$ is the guessing function induced by $\mathfrak{G}^*$ and
\begin{align}
\mathbb{H}^{D, \epsilon}_{\frac{1}{1+\rho}}(X) := \inf_{\substack{P_{\hat{X}|X} : \\
\mathbb{P} [ d(X, \hat{X}) > D ] \leq \epsilon }} H_{\frac{1}{1+\rho}}(\hat{X}).
\end{align}
Also, there exists a guessing strategy with side information $\mathfrak{G}^{*}_{s}$ such that
$P_e (\mathfrak{G}^{*}_{s}) \leq \epsilon$ and
\begin{align}
\frac{1}{\rho} \log \mathbb{E}[\mathfrak{g}^{*} (X | Y)^\rho ]  \leq \mathbb{H}^{D, \epsilon}_{\frac{1}{1+\rho}}(X | Y), 
\end{align}
where $\mathfrak{g}^{*} (\cdot | \cdot)$ is the guessing function induced by $\mathfrak{G}^{*}_{s}$ and
\begin{align}
\mathbb{H}^{D, \epsilon}_{\frac{1}{1+\rho}}(X | Y) := \inf_{\substack{P_{\hat{X}|X,Y} : \\
\mathbb{P} [ d(X, \hat{X}) >  D ] \leq \epsilon }} H_{\frac{1}{1+\rho}}(\hat{X}|Y).   
\end{align}
\end{theorem}

\begin{theorem} \label{ExOneShotConverse}
Fix $D \geq 0$, $\rho > 0$, and $\epsilon \in [0,1)$.
For any guessing strategy $\mathfrak{G}$ satisfying $P_e (\mathfrak{G}) \leq \epsilon$, we have
\begin{align} 
\frac{1}{\rho} \log \mathbb{E}[\mathfrak{g}(X)^\rho ] \geq \mathbb{H}^{D, \epsilon}_{\frac{1}{1+\rho}}(X) - \log \log (1+ \min\{|{\cal X}|, |\hat{\cal X}| \}),
\end{align}
where $\mathfrak{g} (\cdot)$ is the guessing function induced by $\mathfrak{G}$.
Also, for any guessing strategy with side information $\mathfrak{G}_s$ satisfying $P_e (\mathfrak{G}_s) \leq \epsilon$, we have
\begin{align} 
\frac{1}{\rho} & \log \mathbb{E}[\mathfrak{g} (X|Y)^\rho ] \notag \\
& \geq \mathbb{H}^{D, \epsilon}_{\frac{1}{1+\rho}}(X | Y) - \log \log (1+ \min\{|{\cal X}|, |\hat{{\cal X}}| \}),
\end{align}
where $\mathfrak{g} (\cdot | \cdot)$ is the guessing function induced by $\mathfrak{G}_s$.
\end{theorem}

\begin{IEEEproof}
By using the variable-length lossy source code constructed in \cite{Saito18} in lieu of the code in Appendix \ref{code}, we can show Theorem \ref{ExOneShotAchievability} (resp.\ Theorem \ref{ExOneShotConverse}) in the same manner as Theorem \ref{OneShotAchievability} (resp.\ Theorem \ref{OneShotConverse}).
\end{IEEEproof}

Using Theorems \ref{ExOneShotAchievability}, \ref{ExOneShotConverse}, and Eq.\ (46) in \cite{Saito18}, we obtain the next theorem, which shows the asymptotic single-letter characterization of the guessing moment allowing error probability for a stationary memoryless source.
\begin{theorem} 
Suppose that a distortion measure 
$
d_n : {\cal X}^n \times \hat{{\cal X}}^n \rightarrow [0, +\infty)
$
satisfies 
$
d_n(x^n, \hat{x}^n) = \sum_{i=1}^{n} d(x_i, \hat{x}_i)
$
and a source is stationary and memoryless.
Then, for any $\epsilon \in [0,1)$, it holds that
\begin{align} 
\lim_{n \to \infty} \lim_{\rho \downarrow 0} \frac{1}{n\rho} \inf_{\substack{\mathfrak{G}_n }} \log \mathbb{E}[\mathfrak{g}_{n}(X^n)^\rho ] = (1 - \epsilon) R(D, P_X), \label{asymptoticerror}
\end{align}
where the infimum is taken over all guessing strategies $\mathfrak{G}_n$ satisfying $P_e (\mathfrak{G}_n) \leq \epsilon$ and $\mathfrak{g}_{n}(\cdot)$ is a guessing function induced by $\mathfrak{G}_n$.
\end{theorem}

\begin{remark}
Comparing (\ref{asymptotic}) with (\ref{asymptoticerror}), we see that the asymptotic exponent of guessing moment is smaller by a factor of $1 - \epsilon$ by allowing an error probability  at most $\epsilon$.
\end{remark}

\appendix

\subsection{Construction of Variable-Length Lossy Source Code} \label{code}
Let $\{ 0,1 \}^{\star}$ denote the set of all finite-length binary strings 
and the empty string $\lambda$, i.e., 
$
\{ 0,1 \}^{\star} := \{\lambda, 0, 1, 00, 01, 10, 11, 000, \ldots \}.
$
Further, let $w_{i}$ be the $i$-th binary string in $\{ 0,1 \}^{\star}$ in the increasing order of the length and ties are arbitrarily broken.
For example, $w_1 = \lambda, w_2 = 0, w_3 = 1, w_4 = 00, w_5 = 01,$ etc.
Then, we construct the encoder $f$ and the decoder $g$ as follows:

{\bf [Encoder]}
For $x \in {\cal A}_D(\hat{x}^{*}(i)) ~ (i=1, \ldots, \tau)$, 
$
f(x) = w_i.
$

{\bf [Decoder]}
For $i= 1, \ldots, \tau$, 
$
g(w_i) = \hat{x}^{*}(i).
$

From the above construction of the variable-length lossy source code $(f, g)$, it holds that
$
\ell(g^{-1}(\hat{x}^{*}(i))) = \lfloor \log i \rfloor, 
$
where $\ell(\cdot)$ denotes the length of a binary sequence.
Thus, we have
\begin{align} 
\log i \geq \ell(g^{-1}(\hat{x}^{*}(i))). \label{codebound}
\end{align}
Further, from Eq.\ (39) in \cite{Saito18}, for $0 < \beta < 1$, we have
\begin{align} 
H_{\beta}(g(f(X))) = \mathbb{H}^{D}_{\beta}(X). \label{appendfg}
\end{align}

\subsection{Proof of (\ref{majo})} \label{majorization}
To prove (\ref{majo}), we first review the notion of majorization and Schur concave functions.

\begin{defi}
Let $\mathbb{R}_{+}$ be the set of non-negative real numbers 
and $\mathbb{R}^{m}_{+}$ be the $m$-th Cartesian product of $\mathbb{R}_{+}$, where $m$ is a positive integer.
Suppose that ${\bf a} = (a_1, \ldots,$ $a_m) \in \mathbb{R}^{m}_{+}$ and ${\bf b} = (b_1, \ldots, b_m) \in \mathbb{R}^{m}_{+}$ satisfy
$
a_i \geq a_{i+1}, ~ b_i \geq b_{i+1} ~ (i=1,2, \ldots, m-1).
$
If ${\bf a} \in \mathbb{R}^{m}_{+}$ and ${\bf b} \in \mathbb{R}^{m}_{+}$ satisfy, for $k=1, \ldots, m-1$, 
$
\sum_{i=1}^{k} a_i \leq \sum_{i=1}^{k} b_i  
$
and
$
\quad \sum_{i=1}^{m} a_i = \sum_{i=1}^{m} b_i,
$
then we say that ${\bf b}$ {\it majorizes} ${\bf a}$ (it is denoted as ${\bf a} \prec {\bf b}$).
\end{defi}

\begin{defi}
We say that a function $h(\cdot): \mathbb{R}^{m}_{+} \rightarrow \mathbb{R}$ is a {\it Schur concave} function
if $h({\bf b}) \leq h({\bf a})$ for any ${\bf a}, {\bf b} \in \mathbb{R}^{m}_{+}$ satisfying ${\bf a} \prec {\bf b}$.
\end{defi}

\medskip 

Now, we shall show
\begin{align}
\sum_{i=1}^{t} q_i i^{\rho} \geq \sum_{i=1}^{\tau} \mathbb{P} [ X \in {\cal A}_{D} (\hat{x}^{*}(i)) ] i^{\rho}. \label{MajoApp2}
\end{align}
First, we see that $t \geq \tau$ from the definition of $\hat{x}^{*}(i)$ and $q_i$.
Next, we define ${\bf a}, {\bf b} \in \mathbb{R}^{t}_{+}$ as
$
{\bf a} := (q_1, \ldots, q_t)
$,
$
{\bf b} := (\mathbb{P} [ X \in {\cal A}_{D} (\hat{x}^{*}(1)) ], \ldots, \mathbb{P} [ X \in {\cal A}_{D} (\hat{x}^{*}(\tau)) ], \underbrace{0, \ldots, 0}_{t - \tau}).
$
Then, from the definition of $\hat{x}^{*}(i)$ and $q_i$, it holds that ${\bf a} \prec {\bf b}$.
To complete the proof, we consider the function 
$
h({\bf c}) := \sum_{i=1}^{t} c_i i^{\rho},
$
where ${\bf c} = (c_1, \ldots, c_t) \in \mathbb{R}^{t}_{+}$.
Then, from \cite[page 133]{Marshall}, we see that $h(\cdot): \mathbb{R}^{t}_{+} \rightarrow \mathbb{R}$ is a Schur concave function and therefore $h({\bf a}) \geq h({\bf b})$ holds.
Thus, we have (\ref{MajoApp2}).

\end{document}